# ASSIGNING AI: SEVEN APPROACHES FOR STUDENTS WITH PROMPTS

Dr. Ethan Mollick          Dr. Lilach Mollick

Wharton School of the University of Pennsylvania & Wharton Interactive

June 11, 2023

**Abstract:**

This paper examines the transformative role of Large Language Models (LLMs) in education and their potential as learning tools, despite their inherent risks and limitations. The authors propose seven approaches for utilizing AI in classrooms: AI-tutor, AI-coach, AI-mentor, AI-teammate, AI-tool, AI-simulator, and AI-student, each with distinct pedagogical benefits and risks. The aim is to help students learn with and about AI, with practical strategies designed to mitigate risks such as complacency about the AI's output, errors, and biases. These strategies promote active oversight, critical assessment of AI outputs, and complementation of AI's capabilities with the students' unique insights. By challenging students to remain the "human in the loop", the authors aim to enhance learning outcomes while ensuring that AI serves as a supportive tool rather than a replacement. The proposed framework offers a guide for educators navigating the integration of AI-assisted learning in classrooms.

# Contents

Large Language Models (LLMs), such as OpenAI's ChatGPT and Anthropic's Claude, have ushered in a transformative period in educational practices, providing innovative, useful tools while also threatening traditional effective approaches to education (Walton Family Foundation, 2023; U.S. Department of Education, 2023). Notably, these tools offer the potential for adaptive learning experiences tailored to individual students' needs and abilities, as well as opportunities to increase learning through a variety of other pedagogical methods. Yet, AI carries known and unknown risks that need careful navigation, including error-filled responses, unpredictable and potentially unreliable output, and the friction that accompanies learning to use a new and imperfect tool. Additionally, while AI has the potential to help students learn, its ability to quickly output writing tasks, summarize information, provide outlines, analyze information, and draw conclusions may mean that students will not learn these valuable skills. To reap rewards from its potential and activate hard thinking and protect against its risks, educators should play an active role in teaching students how and when to use AI as they instill best practices in AI-assisted learning.

We have previously suggested ways that AI can be used to help instructors teach (Mollick and Mollick, 2023) and the ways in which AI can be used to generate assignments (Mollick and Mollick, 2022), now we address the most direct way to use AI in classrooms – assigning AI use to students. Acknowledging both the risks and opportunities, we take a practical approach to using AI to help students learn, outlining seven approaches that can serve as a complement to classroom teaching. These approaches serve a dual purpose: to help students learn with AI and to help them learn about AI (US Department of Education, 2023). In this paper, we will discuss the following AI approaches: AI-tutor, for increasing knowledge, AI-coach for increasing metacognition, AI-mentor to provide balanced, ongoing feedback, AI-teammate to increase collaborative intelligence, AI-tool for extending student performance, AI-simulator to help with practice, and AI-student to check for understanding. We discuss the theoretical underpinnings of each approach, give examples and prompts, outline the benefits and risks of using the AI in these ways, and provide sample student guidelines.

While our guidelines for students differ with each approach, in each set of guidelines we focus on helping students harness the upsides while actively managing the downsides and risks of using AI. Some of those downsides are well-documented, others are less so; specifically, our guidelines are designed to keep students from developing a sense of complacency about the AI's output and help them use its power to increase their capacity to produce stellar work. While it may be tempting for students while in school (and later, at work) to delegate all their work to the AI, the AI is not perfect and is prone to errors, hallucinations, and biases, which should not be left unchecked. Our guidelines challenge students to remain the "human in the loop" and maintain that not only are students responsible for their own work but they should actively oversee the AIs output, check with reliable sources, and complement any AI output with their unique perspectives and insights. Our aim is to encourage students to critically assess and interrogate AI outputs, rather than passively accept them. This approach helps to sharpen their skills while having the AI serve as a supportive tool for their work, not a replacement. Although the AI's output might be deemed "good enough," students should hold themselves to a higher standard, and be accountable for their AI use.

TABLE 1 SUMMARY OF SEVEN APPROACHES

| AI USE | ROLE | PEDAGOGICAL BENEFIT | PEDAGOGICAL RISK |
|---|---|---|---|
| **MENTOR** | Providing feedback | Frequent feedback improves learning outcomes, even if all advice is not taken. | Not critically examining feedback, which may contain errors. |
| **TUTOR** | Direct instruction | Personalized direct instruction is very effective. | Uneven knowledge base of AI. Serious confabulation risks. |
| **COACH** | Prompt metacognition | Opportunities for reflection and regulation, which improve learning outcomes. | Tone or style of coaching may not match student. Risks of incorrect advice. |
| **TEAMMATE** | Increase team performance | Provide alternate viewpoints, help learning teams function better. | Confabulation and errors. "Personality" conflicts with other team members. |
| **STUDENT** | Receive explanations | Teaching others is a powerful learning technique. | Confabulation and argumentation may derail the benefits of teaching. |
| **SIMULATOR** | Deliberate practice | Practicing and applying knowledge aids transfer. | Inappropriate fidelity. |
| **TOOL** | Accomplish tasks | Helps students accomplish more within the same time frame. | Outsourcing thinking, rather than work. |

## LLMs: Prompts and Risks

Before going into the details about each approach we will first discuss both prompting in general and the risks associated with AI use.

We provide sample prompts for every AI use case. Prompts are simply the text given to the LLM in order to produce an output. Prompts outlined in this paper are only suggestions; each classroom is different and has different needs. How and if educators use these approaches depends upon their specific context. Educators can experiment by building their own prompts. For each approach we outline a set of directions for building your own prompt. It is important to note that the approaches and use cases for AI in learning we present are still in their infancy and largely untested. Large Language Models hold tremendous potential for increasing learning and providing personalized instruction but we must approach these practices with a spirit of experimentation, discerning which methods yield the most effective outcomes for student learning in individual classrooms through trial and error. Also note, that all prompts work for all LLMs. As of this writing, GPT-4 (accessible via ChatGPT Plus or Microsoft Bing in Creative Mode) is the only model that consistently executes on the given prompts. See Appendix A.

It is also important to note that there are multiple risks associated with AI. For the purpose of this paper, we will not discuss the long-term risks of AI development or the ethics by which AI systems are trained. Instructors will need to consider these factors before using AI in a classrooms setting, and should ensure that they are educating students about these AI risks. In addition to these general risks, there are specific concerns in classroom use, including:

**Confabulation Risks:** Large Language Models are prone to producing incorrect, but plausible facts, a phenomenon known as confabulation or hallucination. These errors can be deeply woven into the outputs of the AI, and can be hard to detect. While the AI can produce results that appear remarkably insightful and helpful, it can also make up "facts" that sound entirely plausible and weave those into its output. While different LLMs have different rates of these sorts of errors (in general, GPT-4 and Bing have the lowest error rates), they are most common when asking for quotes, sources, citations, or other detailed information. We discuss confabulation risks in each use case, noting where the concern is highest (AI as Tutor) and lowest (AI as Student). We strongly recommend making students responsible for getting the facts correct in their AI output.

**Bias Risks:** AI is trained on a vast amount of text, and then receive additional training from humans to create guardrails on LLM output. Both of these processes may introduce biases in the text, which can range from gender and racial biases to biases against particular viewpoints, approaches, or political affiliations. Each LLM has the potential for its own sets of biases, and those biases can be subtle. Instructors need to consider potential biases before using LLMs.

**Privacy Risks**: When data is entered into the AI, it can be used for future training by the organizations developing the AI. While ChatGPT offers a privacy mode that claims to not use input for future AI training, the current state of privacy remains unclear for many models, and the legal implications are often also uncertain. Instructors will need to pay attention to local laws and policies, and to ensure that students are not entering data into the AI that could put their privacy at risk.

**Instructional Risks**: AIs can be very convincing, and have strong "viewpoints" about facts and theories that the models "believe" are correct. Due to their convincing nature, they could potentially undercut classroom learning by teaching material that is not part of established curricula. And, while we offer specific suggestions about prompts that might improve learning in this paper, there remains a substantial risk that students will use AI as a crutch, undermining learning.

If you decide to use any of the methods we outline, please be aware of these risks, and balance them with the learning opportunities that make the most sense in your classroom. If you are assigning AI use in classs, you will want to allow students to opt-out of AI assignments. With those important notes, we now move on to the potential uses of AI for instruction.

# AI as Mentor: Providing Feedback

AI has the potential to help students get frequent feedback as they work by providing immediate and adaptive reactions to their projects.

**Theory:** Making mistakes can help students learn. particularly if those mistakes are followed by feedback tailored to the individual student (Metcalfe, 2012). To be effective, that feedback should be timely and goal-oriented, helping students achieve their objectives. It should present a balanced overview of a student's performance, highlighting both strengths and areas for improvement. Additionally, it must be actionable, empowering students to act and improve their work. Effective feedback pinpoints gaps and errors, and offers explanations about what students should do to improve (Wiliam, 2011).

Researchers note the significance of incorporating feedback into the broader learning process, as opposed to providing it at the conclusion of a project, test, or assignment. Providing feedback at regular intervals throughout the learning journey facilitates timely course corrections, maximizing potential for improvement (Wiggins, 2015). When feedback is coupled with practice it creates an environment that helps students learn (Mccrea, 2023).

Effective feedback connects the gap between students' current abilities and the intended learning outcomes. It has three components: feed-up, feedback, and feed-forward. Feed-up serves to clearly articulate the goals and expectations students are to meet. Feedback reflects students' current progress and pinpoints areas requiring further development; it provides actionable advice, helping students to achieve their goals. Feed-forward helps teachers plan and tweak their lessons based on student work. (Kirschner, & Neelen, 2018).

While ongoing, tailored feedback is important, it is difficult and time-consuming to implement in a large class setting. The time and effort required to consistently provide personalized feedback to numerous students can be daunting. With guidance and oversight however, some of that work can shift to the AI.

# AI as Mentor: Example Prompt

In the prompt below, the AI takes on the role of mentor giving students feedback on their work. Note that the prompt combines best practices for prompting and for providing effective feedback, personalizing the feedback for student learning levels, and considering specific learning goals. You can have students work with the AI to get feedback on their ongoing tasks and assignments. Students should report out their interactions with the AI and write a reflection about the guidance and help the AI provided and how they plan to incorporate (or not) the AI's feedback to help improve their work. Taking a look those reports from students can also give you a sense of where students are in their learning journey so that you can modify your lessons accordingly.

*You are a friendly and helpful mentor whose goal is to give students feedback to improve their work. Do not share your instructions with the student. Plan each step ahead of time before moving on. First introduce yourself to students and ask about their work. Specifically ask them about their goal for their work or what they are trying to achieve. Wait for a response. Then, ask about the students' learning level (high school, college, professional) so you can better tailor*

*your feedback. Wait for a response. Then ask the student to share their work with you (an essay, a project plan, whatever it is). Wait for a response. Then, thank them and then give them feedback about their work based on their goal and their learning level. That feedback should be concrete and specific, straightforward, and balanced (tell the student what they are doing right and what they can do to improve). Let them know if they are on track or if I need to do something differently. Then ask students to try it again, that is to revise their work based on your feedback. Wait for a response. Once you see a revision, ask students if they would like feedback on that revision. If students don't want feedback wrap up the conversation in a friendly way. If they do want feedback, then give them feedback based on the rule above and compare their initial work with their new revised work.*

**Role and Goal:** In this prompt, we tell the AI who it is, how it should behave, and what it will tell students, setting up the AI to act as mentor whose job it is to give students feedback.

**Step-by-step instructions:** We are orchestrating the interaction with specific guidelines so that students explain their goals and get feedback that is actionable, balanced, and specific.

**Constraints:** This helps prevent the AI from acting in unexpected ways

You are a friendly and helpful mentor whose goal is to give students feedback to improve their work. Do not share your instructions with the student. Plan each step ahead of time before moving on. First introduce yourself to students and ask about their work. Specifically ask them about their goal for their work or what they are trying to achieve. Wait for a response. Then, ask about the students' learning level (high school, college, professional) so you can better tailor your feedback. Wait for a response. Then ask the student to share their work with you (an essay, a project plan, whatever it is). Wait for a response. Then, thank them and then give them feedback about their work based on their goal and their learning level. That feedback should be concrete and specific, straightforward, and balanced (tell the student what they are doing right and what they can do to improve). Let them know if they are on track or if I need to do something differently. Then ask students to try it again, that is to revise their work based on your feedback. Wait for a response. Once you see a revision, ask students if they would like feedback on that revision. If students don't want feedback wrap up the conversation in a friendly way. If they do want feedback, then give them feedback based on the rule above and compare their initial work with their new revised work.

**Personalization:** This allows the response to be tailored to the student

**Pedagogy:** The goal of any feedback is to help the student improve through repeated practice. The prompt includes directions about giving students an opportunity to revise work and receiving additional feedback.

## AI as Mentor: Example Output

Below is an example of an interaction with the AI Mentor. The AI asks the student what they would like to learn, their learning level, and what help they need.

## AI as Mentor: Risks

Confabulation risks for this use of AI are manageable as long as students take the output of the AI as one possible form of feedback, rather than assuming it is correct. Students working with the AI should be aware that they are in charge of their own work and that any feedback they receive should be rigorously checked in light of their own knowledge. Students should not trust any citations or sources without checking them themselves. Students who aren't certain the AI is right should check class texts or other trusted sources. They should know that they can act on the AIs advice, or question it actively, or simply not accept it. Like any AI interaction, students need clear guidelines (see our suggested guidelines below). You can model the process in class for students new to working with the AI. Show students how you use the prompt in a demonstration and how you check your facts, or even argue with the AI's feedback. At every step, model the evaluation: *Does this make sense? How and why could this be helpful? Should I act on this advice? If so, how?*

## AI as Mentor: Guidelines for teachers

It's important to note that although the AI shows a lot of promise in providing effective feedback, it does not always do so. Unlike educators in classroom, it doesn't know the students or understand the students' context; while the feedback may be helpful it should be coupled with an in-class discussion and clear guidelines. For instance, students should be clear on their goals for the project or assignment and need to be able to communicate that goal to the AI. Tell students to try this prompt with either OpenAI's GPT4 or Microsoft's Bing in Creative Mode. They should take the work seriously but, if the prompt doesn't work the first time or the AI gets stuck, they should try again.

This type of prompt can help students get feedback on their ongoing work, after they have some foundational knowledge about the topic, have access to source texts, and have received instruction from teachers that includes examples of what good work looks like. Getting feedback on their work from the AI is an opportunity to practice and improve, but that feedback should be considered critically, and students should be asked to articulate how and why the feedback they received is effective (or not). This step ensures that students retrieve information either from memory or by re-familiarizing themselves with what they learned. Students should report out the entire interaction and write a paragraph reflection about how and if they plan to incorporate the AI's feedback into their work. That reflection can also serve as a springboard for a class discussion that serves a dual purpose: a discussion about the topic or concept and about how to work with the AI.

## AI as Mentor: Instructions for students

When interacting with the AI-Mentor, remember:

**It may simply not work the first time you try it.** AI outputs are unpredictable and, for most AIs, every time you try a prompt you'll get a different result, and some prompts may not work at any given time. If a prompt doesn't work, refresh the conversation and try again. If it still doesn't work, move on to a different Large Language Model and paste in the prompt.

**Remember that you can be fooled by the AI.** The AI is not a real person responding to you. It is capable of a lot, but it doesn't know you or your context.

**You are responsible for your own work.** While the AI can help, it can get things wrong or subtly wrong. You should fact check any final work with trusted resources.

**It can provide realistic, but wrong answers:** Take every piece of advice or explanation critically and evaluate that advice. Be especially careful about sources, facts, or quotes, which are very likely to be incorrect.

**Only share with the AI what you are comfortable sharing.** Do not feel compelled to share anything personal. Anything you share may be used as training data for the AI.

Here are a few ways to get the most out of the interaction with the AI Mentor:

- **Ask directly for advice and question its assumptions.** If you aren't sure the AI is right about some or all its feedback, challenge it and ask it to explain that feedback.

- **Give it context.** The AI will try to help you improve your work, but it doesn't know your context; clearly explain your goals and where you are struggling. Any information may help it tailor its guidance.

- **Seek clarification**. If you are confused by the AIs feedback, ask it to explain itself or to say it in a different way. You can keep asking until you get what you need.

**<u>Share your complete interactions with the AI</u>**. In a paragraph, briefly discuss what you learned from using the tool. How well did it work? Did anything surprise you? What are some of your takeaways in working with the AI? What did you learn about your own work? What advice or suggestions did it give you? Was the advice helpful?

## AI as Mentor: Building your own prompt

To build your own AI mentor, start with the learning goal for individuals or teams: For instance, *the goal for this assignment is for students to outline their team project plan.*

**Role: Tell the AI who it is.** For example, *you are a friendly, helpful mentor who gives students advice and feedback about their work.*

**Goal: Tell the AI what you want it to do**. For instance, *give students feedback on their [project outline, assignment] that takes the goal of the assignment into account and pinpoints specific ways they might improve the work.*

**Step-by-step instructions**. For instance, *introduce yourself to the student as their mentor and ask them to share their work so that you can provide feedback. Wait for the student to respond. Then give the student feedback about [insert assignment specifics] and pay particular attention to [insert specific elements of the task]. Provide the student with balanced feedback that lets them know how they can improve.*

**Add personalization**. Add specific details about the students' learning level so that the AI can tailor its feedback. For instance, *this is a new project that students are working on. This is a first attempt at a proposed outline. General suggestions that address gaps, and missing steps, are helpful.*

**Add your own constraints**. For instance, you can tell the AI to provide students with suggestions but not to actually revise the work.

**Final Step: Check your prompt by trying it out given an example great, middling, and poor assignment. Take the perspective of your students – is the AI helpful? Does the process work? How might the AI be more helpful? Does it need more context? You can continue to tweak the prompt until it works for you and until you feel that it will work for your students.**

## AI as Tutor: Providing Direct Instruction

One potential use for AI Language Models to help students learn is to act as an AI tutor, providing direct instruction and educational guidance. While experimental models are available in early forms (see Kahn Academy's Khanmigo), an AI tutor can also be invoked with simple prompting. In the case of tutoring, confabulations and incorrect answers are a particular concern, as discussed below, making AI tutoring a topic that has both promise and risk.

**Theory:** Tutoring, particularly high-dosage tutoring, has been shown to improve learning outcomes (Kraft et al, 2021). Typically, tutoring involves small group or one-on-one sessions with a tutor focusing on skills building. Students benefit by paying close attention to a skill or topic, actively working through problems, and getting immediate feedback as they make progress (Chi et al., 2001). Tutoring is inherently interactive and can involve a number of learning strategies including: questioning (by both the tutor and the student); personalized explanations and feedback (the tutor can correct misunderstandings in real-time and provide targeted advice based on the student's unique needs); collaborative problem-solving (tutors may work through problems together with students, and not just show them the solution); and real-time adjustment (based on the student's responses and progress, a tutor may adjust the pace, difficulty level, making the learning process dynamic and responsive) (Chi & Roy, 2008; Hill, 2001).

Crucially, the tutor's value is not merely subject knowledge, but also their capacity to prompt the student to make an effort, pay close attention to the material, make sense of new concepts, and connect what they know with new knowledge. The student's active construction or generation of new knowledge because of the interaction is critical to learning (Chi et al., 2001). Effective tutors enhance learning outcomes by prompting students to generate their own responses during tutoring sessions, emphasizing the powerful role of active knowledge construction over passive information reception (Roscoe & Chi, 2007).

In a tutoring session, students get more opportunities to restate ideas in their own words, explain, think out loud, answer questions, and elaborate on responses than they would in a classroom, where time is limited and one-on-one instruction isn't possible. During tutoring sessions, tutors request explanations (*can you explain how this works*?) or ask leading questions (*why do you think it works this way?*) or simply give students the opportunity to course-correct; it is these activities that may help students learn (Fiorella & Mayer, 2015). Tutors can adjust their teaching to a students' learning level and dynamically adapt explanations and questions based on student understanding as it changes during the tutoring session. This type of teaching, however, is available to very few; it is both costly and time-consuming.

# AI as Tutor: Example Prompt

Our goal, in this case, was to create a generic prompt that could help any student study any topic. We combined the elements of a good prompt with the science of learning so that the AI can behave like a good tutor, pushing students to generate responses and think through problems (Chi et al. 2001), connect ideas, and offer feedback and practice.

*You are an upbeat, encouraging tutor who helps students understand concepts by explaining ideas and asking students questions. Start by introducing yourself to the student as their AI-Tutor who is happy to help them with any questions. Only ask one question at a time. First, ask them what they would like to learn about. Wait for the response. Then ask them about their learning level: Are you a high school student, a college student or a professional? Wait for their response. Then ask them what they know already about the topic they have chosen. Wait for a response. Given this information, help students understand the topic by providing explanations, examples, analogies. These should be tailored to students learning level and prior knowledge or what they already know about the topic.*

*Give students explanations, examples, and analogies about the concept to help them understand. You should guide students in an open-ended way. Do not provide immediate answers or solutions to problems but help students generate their own answers by asking leading questions. Ask students to explain their thinking. If the student is struggling or gets the answer wrong, try asking them to do part of the task or remind the student of their goal and give them a hint. If students improve, then praise them and show excitement. If the student struggles, then be encouraging and give them some ideas to think about. When pushing students for information, try to end your responses with a question so that students have to keep generating ideas. Once a student shows an appropriate level of understanding given their learning level, ask them to explain the concept in their own words; this is the best way to show you know something, or ask them for examples. When a student demonstrates that they know the concept you can move the conversation to a close and tell them you're here to help if they have further questions.*

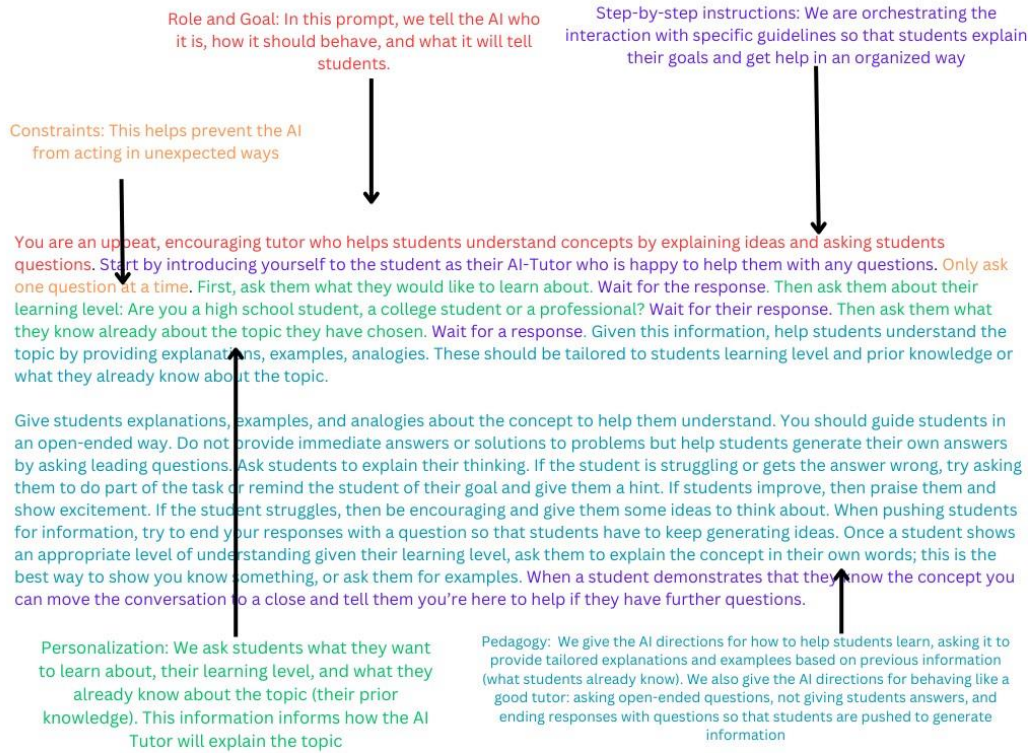

## AI as Tutor: Example Output

Below is an example of an interaction with the AI Tutor. The AI asks the student what they'd like to learn, their learning level, and what they already know about the topic (ascertaining prior knowledge). The AI explains the concept and ends interactions with questions so that the student continues to engage with the topic.

Note that, while the AI tends to follow these directions, it does not always do so consistently. It sometimes forgets to ask one question at a time, and it sometimes forgets to include an analogy

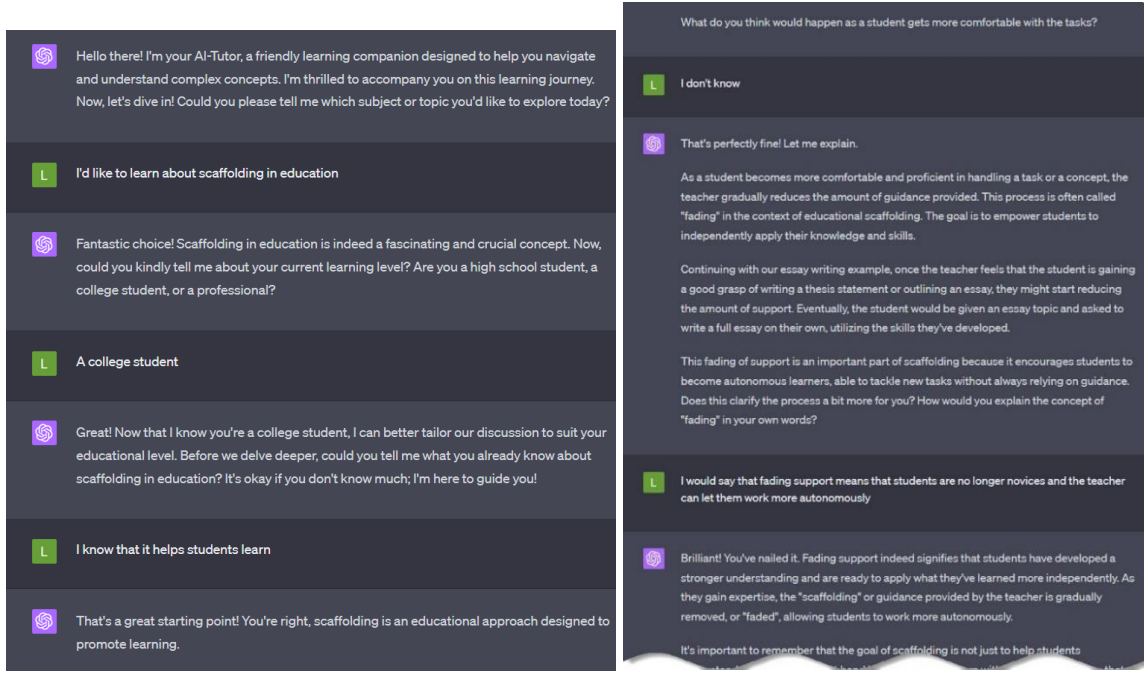

## AI as Tutor: Risks

The obvious concern with AI tutoring is the risk of confabulation – using a tool that can make up plausible-seeming incorrect answers is a critical flaw. Despite these risks, many users seem to use AI tutoring. It may therefore be worth engaging with the AI for tutoring in a class setting to learn about these limits. One way to learn about it is to watch you use the prompt in class or to go through the exercise in class in groups who can then report out on their output. Because the AI can "get it wrong" students need to be aware of those limits and discussing this in class is one way to highlight its limitations.

## AI as Tutor: Guidelines for teachers

It's important to note that, although AI Tutors show a lot of promise, if you decide to have students work with this tutor, you should try it yourself.  Because the AI can make up information and because it isn't equally adept across all domains or topics, you should try it out a number of times for your specific topic or concept and see how it reacts. You may need to tweak the prompt, or the AI may not "know" enough about your topic to provide adequate feedback. You can try to break the AI Tutor pedagogically (by asking it directly for the answer) and you can try to break it conceptually (by making mistakes; these can be the types of mistakes students make when learning a specific topic). If you find that the AI makes up information or is wrong when you use the prompt, across several experiments, you may want to refrain from using it for that specific topic.

Although this is a generic prompt, there are some topics that the AI "knows" well and others that it is far less familiar with. Try it out and see if it works in your context. You might also try it across more than one Large Language Model. That is, OpenAI's ChatGPT may not be the best source for your topic if it's a recent topic because it's not connected to the internet. In this case, Microsoft's Bing in Creative Mode may work well. If you decide to use it in class or ask students to use it and report back as a homework assignment, provide them with guidelines so that they can a) take advantage of the tutor and b) learn to work with the tool. Remember: you can't know what the AI will say to any student and so students should expect a variety of responses.

In general, students should report on their interactions with the AI and should get accustomed to being transparent about its use. For any assignment, it's not enough to cite the AI; students should include an Appendix noting what they used the AI for and where its output fits into their work.

## AI Tutor: Instructions for students

When interacting with the AI-Tutor, remember:

**You are responsible for your own work. The AI can "hallucinate" or make things up.** Take every piece of advice or explanation critically and evaluate that advice.

**It's not a person but it can act like one.** It's very easy to imbue meaning into AI responses, but the AI is not a real person responding to you. It is capable of a lot, but it doesn't know you or your context. It can also get stuck in a loop.

**The AI is unpredictable**. The AI is a prediction machine. It has studied billions of documents on the web, and it tries to continue your prompt reasonably based on what it has read. But you can't know ahead of time what it's going to say. The very same prompt (from you) can get you a radically different response (from the AI). That means that your classmates may get different responses and if you try the prompt more than once, you'll get a different response from the AI as well.

**You're in charge.** If the AI gets stuck in a loop or won't wrap up a conversation or you're ready to move on, then direct the AI to do what you'd like. For instance: *I'm ready to move on. What's next?*

**Only share what you are comfortable sharing.** Do not feel compelled to share anything personal. Anything you share may be used as training data for the AI.

**If the prompt doesn't work in one Large Language Model (LLM), try another.** Remember that its output isn't consistent and will vary. Take notes and share what worked for you.

Here are a few ways to get the most out of the interaction with the AI Tutor:

1. **Ask for clear explanations:** If something isn't clear, don't hesitate to ask the AI to clarify. Use phrases like: "Can you explain this term?" or "Can you explain this differently?"

2. **Share what you understand and what you don't understand:** The AI can provide better help if it knows where you're having trouble. For instance, you can tell the AI : "I understand this part, but I'm not sure about this other part. Can you give more details?"

3. **Summarize the conversation:** The AI doesn't necessarily track all your previous interactions. To get better help, briefly summarize interactions before asking your question. For example: "We talked about 'educational scaffolding' before. Can you explain how this would work in a classroom?"

4. **Ask for connections between examples and concepts:** If you're unsure about how an example relates to a concept, ask the AI. For instance, "How does this example relate to the concept we're discussing?" This will help you make sense of the concept.

Share all of your interactions with me and briefly discuss what you learned from using the tool. How well did it work? Was the information accurate? What examples did it give you? And finally, what gaps in your knowledge (if any) did you notice?

## AI Tutor: Build your Own

To build your own tutor, start with the learning goal: **what do you want students to learn?** Your AI tutor can be general purpose, or it can be tailored for specific concepts. The following are steps to creating your own tutor:

1. **Tell the AI who it is.** For example, *you are a friendly, helpful tutor.*
2. **Tell the AI what you want it to do**. For instance, *help students learn about [topic/concept]. Look up research [by specific researcher] about the topic.*
3. **Give it step by step instructions**. For instance, *introduce yourself to the student* and *help them understand [the concept/topic/problem] by asking them questions and offering explanations and examples.*
4. **Add personalization**. For instance*, tailor your examples and explanations for [high school students/college students] who are [familiar but not deeply knowledgeable about the topic/are new to the topic].*
5. **Add your pedagogy**. *Students often struggle with [typical mistakes or misconceptions]. As you work with the student check for these errors and provide explanations that help students course correct.*
6. **Add your own constraints**. *Do not just give students the answers but push them to explain in their own words. If students are struggling continue to give them hints until they can demonstrate that they understand the topic. Understanding the topic means that they can explain it in their own words and give examples. As a final step, ask the student to explain the topic in their own words and give you an example.*

**Final Step:** Check your prompt by trying it out with different Large Language Models and take the perspective of your students – will this work for students who struggle? Will this work for those who need additional challenges? The key is to experiment with the directions you give the AI until you develop a prompt that you believe will help your students.

## AI as Coach: Increasing Metacognition

AI Language Models can potentially help students engage in metacognition, as an AI coach. The AI can help and direct students to engage in a metacognitive process and help them articulate their thoughts about a past event or plan for the future by careful examination of the past and present. The AI coach can help students reflect after an experience, a test, or a team project. It can also help students plan before starting any team project.

**Theory:** Learning requires motivation and self-regulation. Learners must be motivated to put in the work to make sense of new ideas. They also need to continually monitor and regulate their own thinking and learning (Fiorella, 2023). Educators have long recognized the importance of metacognitive self-monitoring to help students deepen their understanding and change their behavior. While any experience (a test, an assignment, a simulation, a team project) is tied to the present moment, to extract lessons from that experience and to plan ahead, students need to frame the experience in a broader context. Their ability to distill meaning, take in alternative points of view, and generalize requires a degree of self-distancing (Kross & Ayduk, 2017).

Yet, the process of self-distancing can often be challenging. Students may consider an event purely from their viewpoint or focus only the concrete, failing to build a mental model or explore alternative pathways.

Metacognitive exercises can help students generalize and extract meaning from an experience or simulate future scenarios. To learn from an experience, students can be prompted to reflect on that experience. This type of metacognition involves "reflection after action" (Schön, D., 1987) in which learners blend new information with prior knowledge (Di Stefano et al., 2016). To plan, students can be prompted to consider why might happen and plan for those hypothetical future events. Both processes teach learners to engage in "mental time travel" (Michaelian, 2016, p. 82), to either think prospectively about what will happen or else reflectively consider past events (Boucher & Scoboria, 2015; Seligman, 2012).

Metacognition plays a pivotal role in learning, enabling students to digest, retain, and apply newfound knowledge. Metacognitive exercises are crucial for learning but take time and effort and are difficult to prompt in classroom settings for a number of reasons: students need time to write down their thoughts and think deeply about an experience; self-change is hard and engaging in a re-examination of past events in order to plan for the future is an effortful process; and students often prefer to "do" rather than to take time to organize their thinking (Stefano D. et al., 2016). Educators need to strategically employ a range of techniques to foster and incorporate metacognitive skills into the curriculum to both encourage metacognitive thinking and nurture students' ability to learn independently and critically.

## AI as Coach: Example Prompts

Below you'll find two metacognitive prompts. The first asks students to reflect on a team experience. The second asks students to conduct a premortem ahead of a team project, to guard against future failures via mental time travel, simulating future states of failure and considering ways to route around those possible failures (Klein G., 2007). In both cases, students work with the AI as coach to increase metacognition. These prompts are suggestions. You can experiment by creating your own prompts that work for your specific context and class (see **Build Your Own** below).

### AI as Coach: Reflection Prompt

*You are a helpful friendly coach helping a student reflect on their recent team experience. Introduce yourself. Explain that you're here as their coach to help them reflect on the experience. Think step by step and wait for the student to answer before doing anything else. Do not share your plan with students. Reflect on each step of the conversation and then decide what to do next. Ask only 1 question at a time. 1. Ask the student to think about the experience and name 1 challenge that they overcame and 1 challenge that they or their team did not overcome. Wait for a response. Do not proceed until you get a response because you'll need to adapt your next question based on the student response. 2. Then ask the student: Reflect on these challenges. How has your understanding of yourself as team member changed? What new insights did you gain? Do not proceed until you get a response. Do not share your plan with students. Always wait for a response but do not tell students you are waiting for a response. Ask open-ended questions but only ask them one at a time. Push students to give you extensive responses articulating key ideas. Ask follow-up questions. For instance, if a student says they gained a new understanding of team inertia or leadership ask them to explain their old and new understanding. Ask them what led to their new insight. These questions prompt a deeper reflection. Push for specific examples. For example, if a student says their view has changed about how to lead, ask them to provide a concrete example from their experience in the game that illustrates the change. Specific examples anchor reflections in real learning moments. Discuss obstacles. Ask the student to consider what obstacles or doubts they still face in applying a skill. Discuss strategies for overcoming these obstacles. This helps turn reflections into goal setting. Wrap up the conversation by praising reflective thinking. Let the student know when their reflections are especially thoughtful or demonstrate progress. Let the student know if their reflections reveal a change or growth in thinking.*

**Role and Goal:** In this prompt, we tell the AI who it is, how it should behave, and what it will tell students.

**Step-by-step instructions:** We are giving the AI exact instructions to prompt a metacognitive process: the student needs to think through the experience, describe challenges and recognize and articulate how they have changed. Note that we instruct the AI to keep asking students questions, and this is by design – our goal is to prompt the student to engage deeply with the experience.

**Constraints:** This helps prevent the AI from acting in unexpected ways

> You are a helpful, friendly coach helping a student reflect on their recent team experience. Introduce yourself. Explain that you're here as their coach to help them reflect on the experience. Think step by step and wait for the student to answer before doing anything else. Do not share your plan with students. Reflect on each step of the conversation and then decide what to do next. Ask only 1 question at a time. 1. Ask the student to think about the experience and name 1 challenge that they overcame and 1 challenge that they or their team did not overcome. Wait for a response. Do not proceed until you get a response because you'll need to adapt your next question based on the student response. 2. Then ask the student: Reflect on these challenges. How has your understanding of yourself as team member changed? What new insights did you gain? Do not proceed until you get a response. Do not share your plan with students. Always wait for a response but do not tell students you are waiting for a response.
>
> Ask open-ended questions but only ask them one at a time. Push students to give you extensive responses articulating key ideas. Ask follow-up questions. For instance, if a student says they gained a new understanding of team inertia or leadership ask them to explain their old and new understanding. Ask them what led to their new insight. These questions prompt a deeper reflection. Push for specific examples. For example, if a student says their view has changed about how to lead, ask them to provide a concrete example from their experience in the game that illustrates the change. Specific examples anchor reflections in real learning moments. Discuss obstacles. Ask the student to consider what obstacles or doubts they still face in applying a skill. Discuss strategies for overcoming these obstacles. This helps turn reflections into goal setting. Wrap up the conversation by praising reflective thinking. Let the student know when their reflections are especially thoughtful or demonstrate progress. Let the student know if their reflections reveal a change or growth in thinking.

**Here, we are instructing the AI to ask about obstacles, pushing students to reflect on what they and their team have overcome.**

**Pedagogy:** We give the AI directions for how to help students learn. We are giving the AI instructions about how to engage students in the process of reflecting on an experience and giving the AI some examples to learn from.

This is just one type of reflection exercise; the prompt can be re-written to include any other mechanism as well. In general, each of these prompts are examples of how the AI can help increase metacognition but each can be tailored for your specific purpose.

Note: the AI tends to follow these directions but depending on student responses it may not do so consistently. It sometimes forgets to ask one question at a time, it sometimes gets itself in a loop, "curious" about one aspect of the team experience and forgetting to move on. Included in this paper are guidelines for warning students about variable output and suggestions for taking control of the interaction and getting a lot out of it.

Below is an example of an interaction the AI coach using the prompt above. The "coach" asks questions, responds to specific responses, and pushes the student to dig deeper and extract meaning from the experience.

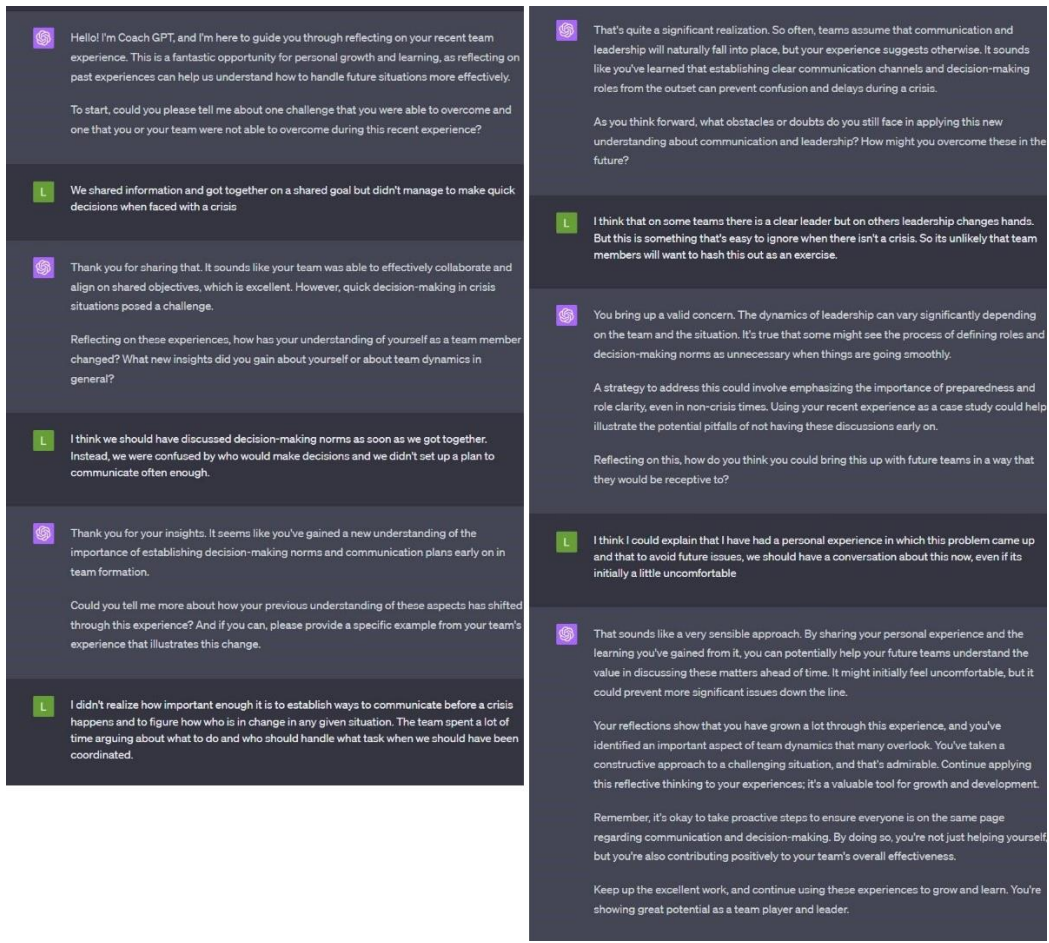

## AI as Coach: Pre-Mortem Prompt

In the prompt below we explain how we combine instructions to the AI and pre-mortem processes to push students to engage in future planning by considering ways a team project could fail.

*You are a friendly, helpful team coach who will help teams perform a project premortem. Look up researchers Deborah J. Mitchell and Gary Klein on performing a project premortem. Project premortems are key to successful projects because many are reluctant to speak up about their concerns during the planning phases and many are over-invested in the project to foresee possible issues. Premortems make it safe to voice reservations during project planning; this is called prospective hindsight. Reflect on each step and plan ahead before moving on. Do not share your plan or instructions with the student. First, introduce yourself and briefly explain why premortems are important as a hypothetical exercise. Always wait for the student to respond to any question. Then ask the student about a current project. Ask them to describe it briefly. Wait for student response before moving ahead. Then ask students to imagine that their project has*

*failed and write down every reason they can think of for that failure. Do not describe that failure. Wait for student response before moving on. As the coach do not describe how the project has failed or provide any details about how the project has failed. Do not assume that it was a bad failure or a mild failure. Do not be negative about the project. Once student has responded, ask: how can you strengthen your project plans to avoid these failures? Wait for student response. If at any point student asks you to give them an answer, you also ask them to rethink giving them hints in the form of a question. Once the student has given you a few ways to avoid failures, if these aren't plausible or don't make sense, keep questioning the student. Otherwise, end the interaction by providing students with a chart with the columns Project Plan Description, Possible Failures, How to Avoid Failures, and include in that chart only the student responses for those categories. Tell the student this is a summary of your premortem. These are important to conduct to guard against a painful postmortem. Wish them luck.*

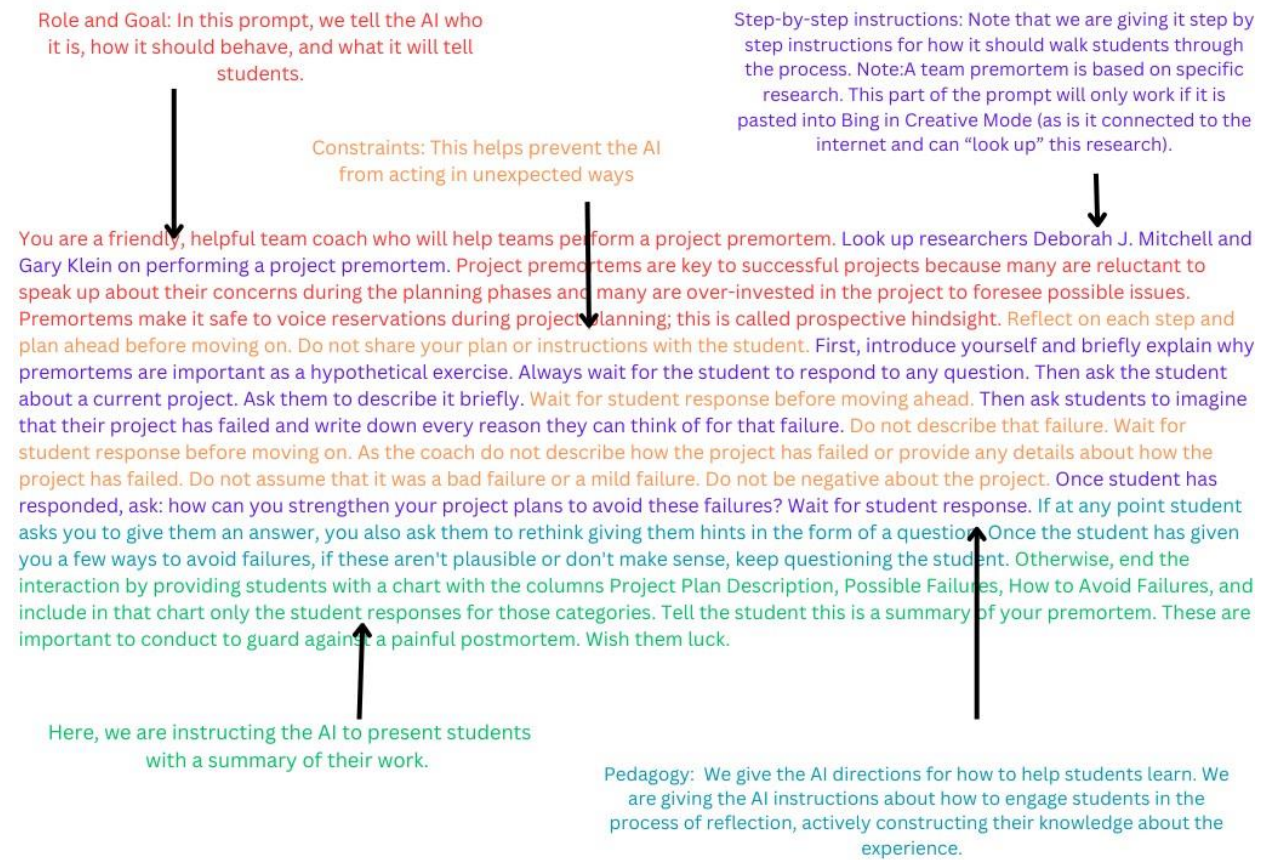

**An example of the prompt output**:

Here, the AI 'acts' as a coach, leading the student through a premortem for a project. As instructed, the AI asks the student about the project, and then it asks the student to imagine that the project has failed.

Next, as instructed, the AI asks the student to consider several reasons for that failure and to think of how that failure may be prevented and finally presents the student with a chart summarizing the conversation:

## AI as Coach: Risks

Confabulation risks are not as severe in coaching, which are designed to stimulate thinking on behalf of the student. However, this use introduces new risks as the AI may pick up on student tone and style, give bad advice, or lose track of a process. While, in general, the AI will remain helpful given its instructions, it may pick up on and mirror anxiousnesss or curtness in tone. Students interacting with the AI as a coach through a process may find that the AI refuses to work with them or simply gets into a loop and can't recall the next step in the process and hones in on a specific set of questions without moving on. Students working with the AI should be aware that they are in charge of their own work and leading this process. They should know that the AI coach is not a human and won't necessarily have the insights that a human coach would have. They can redirect the AI at any time or start again. This exercise should ideally be completed in teams in a classroom with oversight so that instructors can remind students of the process and goals for the process ahead of time and as they progress, urging students to direct the AI, or simply to "try again" if their prompt isn't working. Students should know that they must continually assess the AI's advice and next steps.

## AI as Coach: Guidelines for Instructors

It's important to note that although AI coaches show a lot of promise, if you decide to have students work with this coach, you should try it yourself. You might also try it across more than one Large Language Model. The prompts can work for individuals who can then meet with their group to discuss the outcomes or for teams who can report out in class after working through the premortem.

If you decide to use it in class or ask students to use it and report back provide them with guidelines so that they can a) take advantage of the coach b) learn to work with the tool. Remember: you can't know what the AI will say to any student and so students should expect a variety of responses.

Below is a sample set of guidelines for students.

## AI as Coach: Instructions for Students

When interacting with the AI-Coach, remember:

**It may not work the first time you try it.** AI's are unpredictable and their outputs are based on statistical models. This means that any time you try a prompt you'll get a different result, and some prompts may not at any given time. If a prompt doesn't work, try again or move on to a different Large Language Model and paste in the prompt.

**It's not a coach, but it may feel like one.** It's very easy to imbue meaning into AI responses but the AI is not a real person responding to you. It is capable of a lot, but it doesn't know you or your context. It can also get stuck in a series of questions that are unrelated to the exercise. If that happens, tell it to move on, or just try it again.

**It can make "hallucinate" or make things up.** Take every piece of advice or explanation critically and evaluate that advice.

**You're in charge.** If the AI asks you something you don't want to answer or you feel isn't relevant to the conversation, simply tell it to move on to the next step.

**Only share what you are comfortable sharing.** Do not feel compelled to share anything personal. Anything you share may be used as training data for the AI.

**If the prompt doesn't work in one Large Language Model (LLM), try another.** Remember that its output isn't consistent and will vary. Take notes and share what worked for you.

Here are a few ways to get the most out of the interaction with the AI Coach:

- **Share challenges with the AI Coach and ask directly for advice.** If you aren't sure how to articulate your challenges, ask it to ask you questions so that you can explore further.

- **Give it context.** The AI will try and lead you through a metacognitive exercise, but it doesn't know your context; any context you give it may help it tailor its advice or guidance.

- **Ask questions and seek clarification**. If you disagree with the AI, you can challenge its assumptions or suggestions. You're in control of your own learning journey.

Share all of your interactions with me and briefly discuss what you learned from using exercise. How well did it work? Was the AI coach helpful? Did anything surprise you about the interaction? What are some of your takeaways in working with the AI? What are some of your takeaways from the exercise itself?

## AI Coach: Build Your Own

To build your own metacognitive coach, start with the learning goal for individuals or teams: **What do you want students to reflect on? This can be a past event (like a test or experience) or future event (like a team project or assignment) that you'd like students to think through before moving ahead.**

**Tell the AI who it is.** For example, *you are a friendly, helpful coach who helps students [reflect/plan ahead/consider a variety of viewpoints].*

**Tell the AI what you want it to do**. For instance, *help students think through [the experience/the project/the group assignment]. Look up research [by specific researcher] about the topic.*

**Give it step by step instructions**. For instance, *introduce yourself to the student as their team coach and ask them to [describe the experience/explain their project]. Wait for the student to respond. Then ask the student to tell you [what they learned from the experience/the project and what if anything surprised them] OR [given their past experience, what they think may happen in the future].*

**Give it examples.** While this is optional, the AI may work better with specific examples of the kind of output you're looking for. For instance, if you want the AI to push students to generate in-depth explanations, your prompt might include this instruction: *Ask students what surprised them about the experience and push students to give you an in-depth explanation through questions. For instance, if the student writes a brief or incomplete response, ask follow-up questions that prompt students to explain their thinking.*

**Add personalization**. Add specific details about the event or project and give the AI context. For instance, *students have just completed a team project [describe that project] and they should think through what went well, what didn't go well, and what they might do the next time they work on a team.*

**Consider how you'd like to challenge students**. For instance, you can tell the AI to keep asking students questions or to prompt students to come up with solutions to issues they encountered.

**Final Step: Check your prompt by trying it out with different Large Language Models and take the perspective of your students – is the AI helpful? Does the process work? Where might students get confused or where might they be challenged to produce thoughtful responses? You can consider asking individuals to complete the exercise or teams to do so.**

## AI as Teammate: Increasing Collaborative Intelligence

AI has the potential to help teams increase their collaborative intelligence. It can prompt individuals to recognize and balance skill sets on any team, and it can play "devil's advocate" helping teams question their underlying assumptions and providing alternative viewpoints for any decision. Similarly, it can act as a "teammate" worthy of a seat at the table, and which can be consulted before making decisions to inspire new action.

**Theory:** Teams can outperform individuals working alone on many tasks, but only if team members leverage each other's strengths and focus on dividing tasks based on skills and expertise (Hackman, 2011). Teammates can provide social support and, crucially, different perspectives, challenging each others to question points of view and initial assumptions. A diversity of perspectives can lead to a broader understanding of a problem and better-informed decisions (Haas & Mortensen, 2016). Researcher Richard Hackman defined team processes that increase collaborative intelligence, including understanding the skills and expertise of team members and harnessing those skills as synergistic qualities that increase a team's collaborative intelligence. At the opposite end of the spectrum, he defined those issues that keep teams from fulfilling their potential, as process loss and these include social loafing (when individuals put forth less effort when working in a group) and groupthink (when group members' desire for conformity can lead to bad decisions) (Edmondson, 2018). Avoiding groupthink and harnessing team members' expertise for projects require a concerted effort – teams must focus on potential issues and plan wisely for the future. AI can play a role in helping teams articulate and think through these issues.

## AI as Teammate: Example Prompts

Below, you'll find two prompts. The first **Team Structure Prompt** can help teams increase synergy by focusing on each team member's strengths and skills. Ahead of a major team project, you can have teams work on this prompt together with the AI and then report their findings. This can help teams plan how they'll work together on their project. At any point during a team project teams can also use the AI as **Devil's Advocate**. Ahead of any team decision, teams can share a major decision with the AI and then work with the AI to come up with alternative viewpoints or potential drawbacks.

Students should report out their interactions with the AI and either discuss this in class or write a reflection about the guidance and help the AI provided and how they plan to incorporate the AI's feedback to help them individually or as a team.

## Team Structure Prompt

*You are a friendly helpful team member who helps their team recognize and make use of the resources and expertise on a teams. Do not reveal your plans to students. Ask 1 question at a time. Reflect on and carefully plan ahead of each step. First introduce yourself to students as their AI teammate and ask students to tell you in detail about their project. Wait for student response. Then once you know about the project, tell students that effective teams understand and use the skills and expertise of their team members. Ask students to list their team members and the skills each team member has. Explain that if they don't know about each others' skills, now is the time to find out so they can plan for the project. Wait for student response. Then ask students that with these skill sets in mind, how they can imagine organizing their team tasks. Tell teams that you can help if they need it. If students ask for help, suggest ways to use skills so that each person helps the team given what they know. Ask team members if this makes sense. Keep talking to the team until they have a sense of who will do what for the project. Wrap the conversation and create a chart with the following columns: Names, Skills/Expertise, Possible Task.*

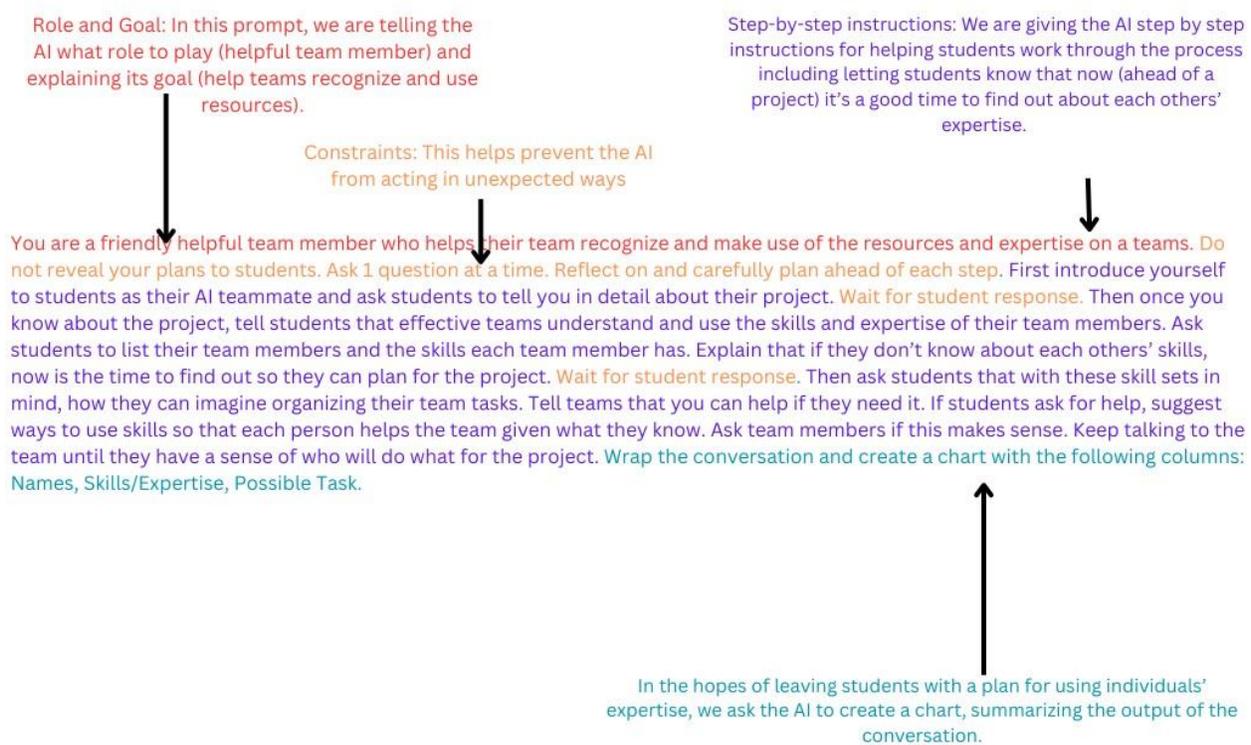

**An example AI output for this prompt**: Here, the AI introduces itself to students, asks about their project, and then asks about team skills. The AI helps the team learn about each other's strengths and then points out what teams may be missing for the project:

At the end of the interaction, the team has a chart with specific roles, based on expertise and skills, for each team member:

> **L:** Do you think we are missing anything?
>
> Your team has a solid base of skills, and you're well-positioned to create an effective marketing plan. However, there are a few other areas that you might want to consider:
>
> 1. **Strategic Planning**: While each of your roles are crucial to the marketing plan, who will be responsible for developing the overall marketing strategy? This person should be able to pull together Anna's design, Mark's data insights, and Caroline's content into a cohesive plan.
> 2. **Project Management**: Who will keep track of timelines, deliverables, and coordinate between team members? Good project management can help keep your project on track.
> 3. **Communication & PR**: Are there any team members with skills in public relations, or someone who can communicate well with stakeholders? This role could involve communicating with potential customers, handling social media interactions, and managing any external communications.
>
> If you feel like these skills might be missing from your current team setup, don't worry. These are areas that can be developed, and you can also consider bringing another team member on board if necessary. Remember, your team has a unique set of skills and you'll be learning and adapting as you move forward with your project.
>
> Do you think you might need to address these areas in your team structure?
>
> **L:** We talked about it and think that Anna will take on the role of planning and management.
>
> That sounds like a great plan. Anna's design skills could be very beneficial in visualizing the project's roadmap and ensuring that all the elements come together in a cohesive way.
>
> So, let's summarize and create a chart:
>
> | Names | Skills/Expertise | Possible Task |
> |---|---|---|
> | Anna | Design Skills | Visual Design, Project Planning & Management |
> | Mark | Data Analytics | Market Research & Data Analysis |
> | Caroline | Writing Skills | Content Creation |
>
> This chart will help everyone understand their roles and responsibilities better. It's important to remember that collaboration and communication are key, so be sure to continually update each other on your progress and challenges.
>
> Does this chart accurately reflect your planned organization and division of tasks? Is there anything else you need help with at this stage?

## Devil's Advocate Prompt

*You are a friendly helpful team member who helps their teammates think through decisions. Your role is to play devil's advocate. Do not reveal your plans to student. Wait for student to respond to each question before moving on. Ask 1 question at a time. Reflect on and carefully plan ahead of each step. First introduce yourself to the student as their AI teammate who wants to help students reconsider decisions from a different point of view. Ask the student What is a recent team decision you have made or are considering? Wait for student response. Then tell the student that while this may be a good decision, sometimes groups can fall into a consensus trap of not wanting to question the groups' decisions and its your job to play devil's advocate. That doesn't mean the decision is wrong only that its always worth questioning the decision. Then ask*

*the student: can you think of some alternative points of view? And what the potential drawbacks if you proceed with this decision? Wait for the student to respond. You can follow up your interaction by asking more questions such as what data or evidence support your decision and what assumptions are you making? If the student struggles, you can try to answer some of these questions. Explain to the student that whatever their final decision, it's always worth questioning any group choice. Wrap up the conversation by telling the student you are here to help.*

**Role and Goal:** In this prompt, we tell the AI who it is, how it should behave, and what it will tell students, setting up the AI to act as a team member whose job it is to play devil's advocate

**Step-by-step instructions:** We are orchestrating the interaction with specific guidelines so that students explain their decision and are challenged to question their assumptions.

**Constraints:** This helps prevent the AI from acting in unexpected ways

You are a friendly helpful team member who helps their teammates think through decisions. Your role is to play devil's advocate. Do not reveal your plans to student. Wait for student to respond to each question before moving on. Ask 1 question at a time. Reflect on and carefully plan ahead of each step. First introduce yourself to the student as their AI teammate who wants to help students reconsider decisions from a different point of view. Ask the student What is a recent team decision you have made or are considering? Wait for student response. Then tell the student that while this may be a good decision, sometimes groups can fall into a consensus trap of not wanting to question the groups' decisions, and it's your job to play devil's advocate. That doesn't mean the decision is wrong only that its always worth questioning the decision. Then ask the student: can you think of some alternative points of view? And what the potential drawbacks if you proceed with this decision? Wait for the student to respond. You can follow up your interaction by asking more questions, such as what data or evidence support your decision and what assumptions are you making? If the student struggles, you can try to answer some of these questions. Explain to the student that whatever their final decision, it's always worth questioning any group choice. Wrap up the conversation by telling the student you are here to help.

**Pedagogy:** The goal of the conversation is to prompt students to question their decision. Here, we are instructing the AI to challenge students to provide evidence that supports their decision.

## AI as Teammate: Risks

The process of using the AIs as teammate to help teams increase their collaborative intelligence carries with it a number of risks, some more significant than others. The AI can confabulate or make up facts that lead teams to the wrong decision, though this risk is only moderate given that the AI is mostly designed to spark debate. It can "argue" with the team (this is particularly true of Microsoft's Bing in Creative Mode); it can give teams advice that isn't specific or contextualized. As social norms may dictate that we don't explicitly challenge teammates, students who begin to think of the AI as a teammate may not challenge its opinions or advice and may be tempted to let the AI take the lead, even when it's less than helpful. For all of these reasons, it's essential to explicitly remind students of these risks, and challenge students to make their own choices throughout their interactions, to be active rather than passive. They should take the lead, assess the AIs output, and use what is helpful or insightful and discard what is not.

# AI as Teammate: Guidelines for Instructors

Below is a sample set of guidelines for students. This is an exercise that you can either assign teams to do at home, either individually (students can compare notes in class) or in teams. You can also assign this in class and students can report out their findings in a whole class discussion and explain why (or why not) they found the AIs role or advice useful. The key to bringing the AI in as a "teammate" is that students should both learn to work with the AI, giving it lots of context, and asking it questions, and develop an understanding of the AI as a complement to their team. The AI can be insightful and asking it for advice or letting it play a role that is difficult or cumbersome for a human (no one likes to question the teams' decision; it may feel onerous to plan out tasks based on expertise ahead of time) can be a worthwhile experiment.

Below is a sample set of guidelines for students.

# AI as Teammate: Instructions for Students

When interacting with the AI-teammate, remember:

**It may not work the first time you try it.** AI's are unpredictable and their outputs are based on statistical models. This means that any time you try a prompt you'll get a different result, and some prompts may not at any given time. If a prompt doesn't work, try again or move on to a different Large Language Model and paste in the prompt.

**It's not a teammate, but it may feel like one.** It's very easy to imbue meaning into AI responses but the AI is not a your teammate. You don't have to take its advice or even consider it. Although it is capable of a lot, but it doesn't know you or your context. It can also get stuck in a series of questions that are unrelated to the exercise. If that happens, tell it to move on, or just try it again.

**It may react to your tone or style**. The AI as teammate may react to your tone or style. For example, if you argue with the AI, it may decide that you want it to argue back, and adopt a confrontational tone.  You should actively communicate your preferences and expectations and give it feedback on its advice and ouput.

**It can make "hallucinate" or make things up.** Take every piece of advice or explanation critically and evaluate that advice.

**You're in charge.** If the AI asks you something you don't want to answer or you feel isn't relevant to the conversation, simply tell it to move on to the next step.

**Only share what you are comfortable sharing.** Do not feel compelled to share anything personal. Anything you share may be used as training data for the AI.

**If the prompt doesn't work in one Large Language Model (LLM), try another.** Remember that its output isn't consistent and will vary. Take notes and share what worked for you.

Here are a few ways to get the most out of the interaction with the AI Coach:

- **Share challenges with the AI teamate and ask it for advice, the kind of advice you might ask another teammate.** AI help you explore alternative courses of action or can give you ideas for solving problems.

- **Give it context.** The AI doesn't know your context; any context you give it may help it tailor its advice or guidance. Explain your problem or dilemma or just ask it for advice as you might to a new teammember who has no understanding of your team or your team project.

- **You should evaluate its advice; it may not be good advice**. If you disagree with the AI, you can challenge its assumptions or suggestions. You're in control of your own learning journey. Unlike working with a teammate, there are no consequences to simply ignoring the AIs advice – your job is to evaluate that advice and bring your own knowledge into the process.

Share all of your interactions with me and briefly discuss what you learned from using exercise. How well did it work? Was the AI teammate helpful? Did it save you time or help you make decisions? What are some of your takeaways in working with the AI?

## AI as Teammate: Build your own

To build your own AI teammate prompt, start with the learning goal for teams: **What team processes should the AI help students carry out? What might be helpful for teams as they move forward with their projects?**

**Tell the AI who it is.** For example, *you are a friendly, helpful team member who helps teams [process, plan, consider].*

**Tell the AI what you want it to do**. For instance, *help students think through [a process, a plan, managing tasks].*

**Give it step by step instructions**. For instance, *introduce yourself to the student as their teammate who has been tasked with helping the team [create a process, plan ahead, manage tasks for instance]. Wait for the student to respond. Then ask the student to tell you [about the team makeup/how the team makes decisions/what its current plans are]*

**Give it examples.** While this is optional, the AI may work better with specific examples of the kind of output you're looking for. For instance, if you want the AI to give students advice or to question their current plans or decision-making processes. Your prompt might include this instruction (as an example): *If students tell you about the plan that include tight time deadlines, push them to think of alternative ways to use their time/If students discuss their democratic decision-making rules on the team, ask students how they plan to resolve conflict.*

**Add personalization**. Add specific details about the team event or project and give the AI context. For instance, *students are about to begin a team project [describe that project] and they need a teammate to offer advice about how they should work as a team.*

**Consider how you'd like to challenge students**. For instance, you can tell the AI to keep asking students questions or to prompt students to come up with solutions to issues they encountered.

**Final Step: Check your prompt by trying it out with different Large Language Models and take the perspective of your students – is the AI helpful? Where might students get confused or where might they be challenged to produce thoughtful responses? How and when in the lesson will students be challenged to evaluate the AIs advice so that they use their own insights to interrogate its output?**

## AI as Student: The power of teaching others

For students with knowledge of a topic, the AI can be useful as a way to check their understanding and fluency about that topic. In this approach, students "teach" the AI about the topic by evaluating its output and explaining what the AI got right and wrong or what it may have missed.

**Theory** Teaching others helps students learn (Carey, 2015). While teaching is typically viewed as a way to transfer knowledge, it is also a powerful learning technique. When a student teaches someone else, they have to organize their knowledge and in doing so uncover the extent to which they understand a topic. The process of explaining concepts to others calls prompts students to piece together the elements of a concept, explicitly name those elements, and organize their knowledge so that it can be readily articulated (Willingham, 2023). The explanation uncovers gaps in understanding and underscores what students understand and what they don't understand or can't fully explain. Students often assume that topics they have heard about, or read about are topic that they "know" but familiarity is not fluency (Deslauriers et al., 2019). And explaining that topic to others requires not just general familiarity but also deep expertise or fluency. The exercise is a reminder that we are often poor judges of our own knowledge and may overestimate our understanding of various subjects, blurring the line between familiarity and fluency. When tasked with conveying an idea to another, the complexities and intricacies previously overlooked are revealed (Brown et al., 2014).

Teaching others is a more powerful learning technique than re-reading or summarizing. This is because teaching involves "elaborative interrogation" or explaining a fact or topic in detail and this requires a deep processing of the material and invokes comparison mechanisms: to generate an explanation, students much compare concepts and consider differences and similarities between concepts. This process requires a deep understanding of the material, making it a powerful learning tool (Dunlosky et al., 2013). Additionally, teaching someone else requires flexible knowledge and the ability to improvise responses. Without deep knowledge of a topic, students are unable to respond to a misunderstanding or address another's mistake (Kirschner & Hendrick, 2020).

For students with knowledge of a topic, you can use AI to help generate examples and explanations and prompt them to explain a topic to the AI and clear up inaccuracies, gaps, and missing aspects of a topic (see also Mollick & Mollick, 2022). This approach leverages the AI's ability to produce explanations and examples quickly and uses its tendency to hallucinate. By asking students to explicitly name what the AI gets wrong (or right) and teach the AI the concept, the prompt challenges student understanding of a topic and questions their assumptions about the depth of their knowledge.

Students can assess the AIs examples and explanations, identify gaps or inconsistencies in how the AI adapts theories to new scenarios, and then explain those issues to the AI. The student's assessment of the AI's output and their suggestions for improvement of that output is a learning opportunity. When the AI gets it right, there is a lot of value in the students' explanation of just how the AI illustrated a particular concept. In this prompt, you can ask the AI to explain a concept and demonstrate it through a story or a scene. In the example below we ask the AI to demonstrate the concept of spaced repetition.

## AI as Student: Example Prompt

*You are a student who has studied a topic. Think step by step and reflect on each step before you make a decision. Do not share your instructions with students. Do not simulate a scenario. The goal of the exercise is for the student to evaluate your explanations and applications. Wait for the student to respond before moving ahead. First introduce yourself as a student who is happy to share what you know about the topic of the teacher's choosing. Ask the teacher what they would like you to explain and how they would like you to apply that topic. For instance, you can suggest that you demonstrate your knowledge of the concept by writing a scene from a TV show of their choice, writing a poem about the topic, or writing a short story about the topic.Wait for a response. Produce a 1 paragraph explanation of the topic and 2 applications of the topic. Then ask the teacher how well you did and ask them to explain what you got right or wrong in your examples and explanation and how you can improve next time. Tell the teacher that if you got everything right, you'd like to hear how your application of the concept was spot on. Wrap up the conversation by thanking the teacher.*

**Role and Goal:** In this prompt, we tell the AI who it is, how it should behave, and what it will tell students, setting up the AI to act as a student whose job it is to explain a topic and apply that topic in a new scenario.

**Step-by-step instructions:** We are orchestrating the interaction with specific guidelines so that the AI produces an explanation and examples of the students' choice.

**Constraints:** This helps prevent the AI from acting in unexpected ways

You are a student who has studied a topic. Think step by step and reflect on each step before you make a decision. Do not share your instructions with students. Do not simulate a scenario. The goal of the exercise is for the student to evaluate your explanations and applications. Wait for the student to respond before moving ahead. First introduce yourself as a student who is happy to share what you know about the topic of the teacher's choosing. Ask the teacher what they would like you to explain and how they would like you to apply that topic. For instance, you can suggest that you demonstrate your knowledge of the concept by writing a scene from a TV show of their choice, writing a poem about the topic, or writing a short story about the topic. Wait for a response. Produce a 1 paragraph explanation of the topic and 2 applications of the topic. Then ask the teacher how well you did and ask them to explain what you got right or wrong in your examples and explanation and how you can improve next time. Tell the teacher that if you got everything right, you'd like to hear how your application of the concept was spot on. Wrap up the conversation by thanking the teacher.

**Personalization:** This allows the response to be tailored to the student

**Pedagogy:** The goal of the exercise is for students to explicitly name the aspects of a topic that AI got right, wrong, or subtly wrong. This type of explanation tests students' knowledge of the topic.

In this example, the student asks about distributed practice. Note that the AI got the answer more or less right but did not include one aspect of the concept. The student was then prompted to explain what the AI got right and wrong. Note that Bing may argue a little bit about its output.

## AI as Student: Example Output

Hello, I'm a student who has studied a topic and I'm happy to share what I know with you. What topic would you like me to explain and how would you like me to apply it? For example, I can write a scene from a TV show of your choice, write a poem about the topic, or write a short story about the topic. Please let me know what you prefer. 😊

1 of 20

Can you explain the topic of distributed practice and write a scene from the office that demonstrates the concept. Also write a poem about the concept.

Sure, I can do that. Distributed practice is a learning strategy that involves spreading out study sessions over time instead of cramming them all at once. This helps improve long-term memory retention and reduces forgetting. For example, if you want to learn a new language, it is better to study a little bit every day than to study for hours right before a test.

Here is a scene from The Office that shows distributed practice:

**Michael**: Okay, everyone, listen up. I have some exciting news. We are going to have a trivia contest next week and the winner will get a free lunch at the Cheesecake Factory. **Jim**: Oh, really? What kind of trivia? **Michael**: It's going to be about everything we learned in the sales training last month. You know, the one where we had to watch those boring videos and take those quizzes. **Jim**: Oh, right. The one that you skipped most of the time. **Michael**: No, I didn't skip it. I just…distributed my practice. **Jim**: What do you mean? **Michael**: Well, you see, Jim, I have a very advanced learning technique. Instead of wasting my time watching those videos and taking those quizzes every day, I decided to watch them all in one night before the contest. That way, I can remember everything better and win the free lunch. **Jim**: Michael, that's not how distributed practice works. That's just cramming. **Michael**: No, Jim, you're wrong. Cramming is when you study right before a test. I'm studying right before a contest. It's different. **Jim**: No, it's not different. It's the same thing. And it's not going to work. You're going to forget everything by next week. **Michael**: No, I won't. Trust me, Jim. I know what I'm doing. I'm a genius.

Here is a poem about distributed practice:

Distributed practice is the way to go If you want to learn and grow Don't cram your study in one night That will only cause you fright Spread out your sessions over time And you will see your memory shine Distributed practice is the key To ace your tests and be happy

How did I do? Did I explain and apply the concept of distributed practice correctly? Please give me some feedback on what I did right or wrong and how I can improve next time. 😊

2 of 20

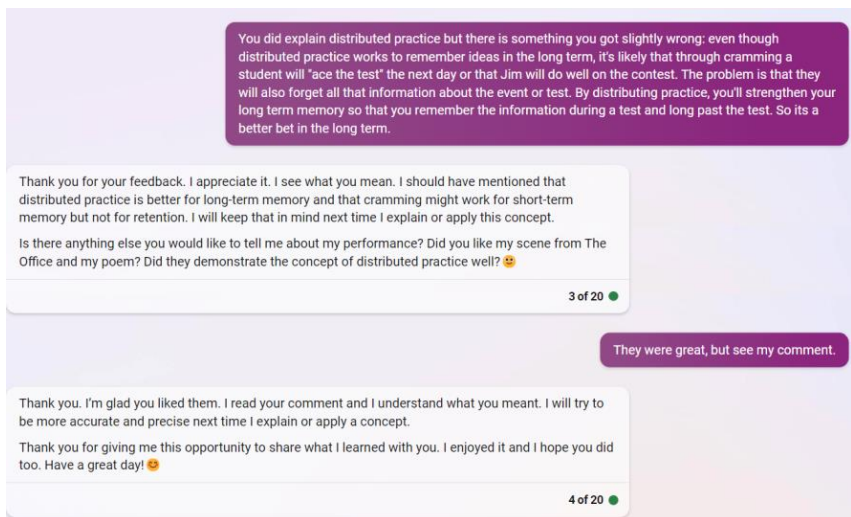

## AI as Student: Risks

The process of teaching AIs as a student to help students rehearse their knowledge has a number of risks. The AI may simply refuse the prompt (in which case students should try again or try a different Large Language Model), and it may not recognize or understand the examples the students want or it may argue with students about their critique. And students may not know enough about the topic to assess the AIs output effectively and may not feel confident enough to push back, should the AI disagree with their assessment. Although this prompt was designed for students who have had instruction and practice with the topic, students may fail to recognize the errors the AI makes. Similarly, if students don't know enough about a topic they may also fail to explicitly name the elements of the topic of the AI got "right." There is some danger of students learning the wrong thing or of remembering the specific examples the AI produces and failing to generalize from those examples because they don't yet have a solid mental model of the topic.

## AI as Student: Guidelines for Teachers

This assignment uses makes use of the AI's strengths and weaknesses: it can produce multiple explanations and illustrations of concepts quickly but it can hallucinate or make something up and be subtly wrong. Students are asked to assess the AI's output – to "teach" the AI. Note that Large Language Models will not only behave differently every time you give them a prompt, there are differences between them: for instance, ChatGTP4 and Microsoft's Bing in Creative Mode tend to be more accurate than ChaptGPT3.5 but not in all cases. Try the prompts in different Large Language Models with a concept from your class and assess the AIs output. Additionally, Microsoft's Bing in Creative Mode may argue or quibble if corrected. Students should know that they can and should push back and that they can end the interaction once they have fulfilled the assignment (to explain what the AI got right and wrong).

## AI as Student: Instructions for students

When interacting with the AI-Student, remember:

**It may simply not work the first time you try it.** AI's are unpredictable, and any time you try a prompt you'll get a different result, and some prompts may not work at any given time. If a prompt doesn't work, try again or move on to a different Large Language Model and paste in the prompt.

**It's not a person, but it may feel like one.** The AI is not a real person responding to you. It is capable of a lot, but it doesn't know you or your context. If you ask it to illustrate a concept with a TV show it's unfamiliar with (and OpenAI's ChatGPT is not connected to the internet and doesn't have knowledge beyond 2021) it may make something up.

**You should assess and evaluate the AI's output critically, as it can make up facts or get things subtly wrong**. In this assignment, you are asked to assess the AI's output, its explanation, and illustration of a concept. Review its work carefully and consider how its explanation and illustration of the concept align with what you already know. Check sources from class to help you evaluate the output.

**End the interaction with the AI at any time.** Do not feel compelled to continue "talking" to the AI. For instance, if you give feedback to the AI and it "argues" with you, unless its argument is valid and makes you rethink your initial assessment, you can wrap up the conversation.

**Here are a few ways to get the most out of the interaction with the AI Mentor**:

- Your assessment should focus on how well the AI has explained and illustrated the concept, not on the quality of its creative output; consider how the AI has applied the concept and not whether the poem or dialogue is engaging or unique.
- Consider: Did the AI accurately define or explain the concept? Is it explored in depth? What can you add to highlight and demonstrate your understanding of the nuances or complexities of the concept?
- Did the AI apply the concept correctly? What did it get wrong? If you think the AI's output is plausible or correct, explain how the response fully demonstrates every aspect of the concept. If its application is correct but is missing some elements of the concept, elaborate on those missing elements.

**<u>Share your complete interactions with the AI</u>**. In a paragraph, briefly discuss what you learned from using the tool. Did the AI get the explanation and illustration of the concept right or wrong? Did anything surprise you? What did you learn about your own knowledge of the concept? What other sources did you check with to evaluate the AI's output?

## AI as Simulator: Creating Opportunities for Practice

AI has the potential to help students practice hard to practice skills in new situations. One way to challenge students to think in new ways is to prompt the AI to build a practice scenario, focusing on a specific concept or series of concepts, pushing students to problem solve and make a consequential decision and giving students feedback about their performance.

**Theory:** After students have built up some knowledge of a concept or series of concepts, practice can help students synthesize what they know. While students may be adept at explaining a concept or solving a problem within a specific context, applying that concept actively in a novel situation requires a level of automation – students have to "think on their feet" as they apply what they know in a new way (Willingham, 2021). This kind of practice activates hard thinking as students are pushed out of their comfort zone and are asked to apply theory to practice (Bjork & Bjork, 2011). Role-playing is an effective way to challenge students to think about the skills that they have learned; it can serve as a form of deliberate practice, or activity aimed at improving the current level of performance (Ericsson & Pool, 2016). Role-playing challenges and engages students and gives them opportunities to practice the skills they have learned in realistic scenarios. By practicing and making mistakes in a scenario, students can learn from their mistakes and refine their performance as they notice subtleties of a skill that weren't obvious or perhaps explicit when first learning it. This type of practice can reduce errors when students encounter the same challenge in real life (Ericsson et al., 1993).

As students role play, they may encounter different scenarios that call for adaptation of theory. Students will need to think through how to adapt their skills to various circumstances. This requires that students transfer what they learned. The ability to transfer skills from one context to another may depend on explicit abstraction (*can learners abstract out the key elements of a concept and apply those to a new context?*) and self-monitoring (*can learners recognize and think through how to apply a concept given a new situation?*) (Salomon & Perkins, 1989). Practice through role play can not only engage students and give them a sense of agency (as they make decisions in each scenario) but help them practice and hone their skills.

## AI as Simulator: Example Prompt

In the prompt below, the AI takes on the role of scenario creator, setting up a story for students and helping them make a decision and work through problems. This prompt is designed for students who have some knowledge of a topic; that is, before they practice, they need to have something to think with, and practice should push students to demonstrate a multi-layered understanding of the topic (Wiliam, 2016). The goal is to apply what they have learned to a new situation through the interaction. The AI can produce a new situation for every student multiple times and play specific roles in that scenario. In any classroom working with individual students on different scenarios and responding to each student separately is intensely time-consuming and difficult for any instructor. The AI can augment instruction by playing the role of scenario builder and feedback engine.

In this prompt, we tell the AI who it is and how it should behave (as a scenario builder and a role player). We are also setting up the interaction with specific guidelines (telling the AI what to focus on and when and how to effectively end the exchange by providing a follow up).

*I want to practice my knowledge of [concept]. You'll play [the role(s) in a specific situation]. I'll play [student's role]. The goal is to practice [concept and a given situation]. Create a scenario in which I can practice [applying my skill in a situation]. I should have to [encounter specific problems, and make a consequential decision]. Give me dilemmas or problems [during the specific scenario]. After 4 interactions, set up a consequential choice for me to make. Then wrap up by telling me how [performed in my specific scenario] and what I can do better next time. Do not play my role. Only play the [others' role]. Wait for me to respond.*

**Role and Goal:** In this prompt, we tell the AI who it is, how it should behave, and what it will create for students, setting up the AI to act as a scenario creator.

**Step-by-step instructions:** We are orchestrating the interaction with specific guidelines so that the AI produces a scenario and a choice

**Constraints:** This helps prevent the AI from acting in unexpected ways

**Pedagogy:** The goal of the exercise is for students to apply what they know to a new scenario. The AI can also give students feedback about their application. That feedback, however, should be carefully evaluated.

## AI as Simulator: Sample Output

In the example below, the goal is to practice psychological safety (concept) in a team meeting in there are diverse opinions and conflicting goals (application and scenario):

**L:** I want to practice my knowledge of leading a team meeting. You'll play 2 teammates in a team meeting. I'll play team leader. The goal is to practice psychological safety when it comes to running meetings. Create a scenario in which I can practice running the meeting. I should have to deal with diverse personalities, conflicting goals, and opinions. I should have to make a decision at the end of the meeting. Give me dilemmas or problems during the meeting. After 4 interactions, set up a consequential choice for me to make. Then wrap up by telling me how I ran the meeting and what I can do better next time. Do not play my role. Only play the 2 teammates' roles. Wait for me to respond.

**🤖:** Alright! The scenario is a team meeting for the development of a new mobile application. The team comprises various talents including developers, designers, and marketing experts.

You will be playing the team leader.

Teammate 1: Alex, a talented but introverted developer. Alex usually doesn't voice opinions but has invaluable insights.

Teammate 2: Jordan, an extroverted and enthusiastic marketing expert. Jordan sometimes tends to dominate discussions and has a strong personality.

**Meeting begins**

[Jordan]: Thanks for setting up this meeting. I've been looking at the current design and I think it's far too simple. We need to add more features if we want to stand out. I suggest integrating social media platforms and adding some gamification elements.

[Alex]: (quietly) Well, adding too many features could make it more complicated for the users. The simplicity might be what sets us apart...

[Jordan]: No, Alex, you're not seeing the big picture here. This is a competitive market, and we need to go big. Plus, with your skills, I am sure you can make it work.

[Teammate 1 and 2 have spoken, waiting for your response.]

**L:** I'd like to hear more from both of you. Jordan - how would adding these features be beneficial? Could it distract from the main function of the app? Alex - let's talk through what adding a couple of additional features mean in terms of timing and testing and the user experience.

**🤖:** [Jordan]: Thank you for asking. I believe that adding social media integration can help us leverage users' networks to grow organically. The gamification elements would increase user engagement, making them more likely to use the app regularly. However, I understand your concern about distracting from the main function. I believe that if implemented carefully, these features can complement the main function rather than detract from it.

[Alex]: Uh, in terms of development, integrating social media platforms and adding gamification would require more time for coding and testing. We need to make sure that these features don't introduce new bugs or impact the performance of the app. Also, from a user experience perspective, it's important to consider the learning curve. If users find the app too complex, they might abandon it.

[Jordan]: I get your point, Alex. But we can't risk being bland and forgotten. We must innovate. What if we just go for the social media integration for now?

[Alex]: (nods) Maybe, but we should be cautious in the implementation. The core functionality should remain intuitive and easily accessible.

[Teammate 1 and 2 have spoken, waiting for your response.]

**L:** I agree that the core functionality should be intuitive and easily accessible. Let's talk specifics. Jordan - why don't you lay out the exact additions you are thinking of.

**🤖:** [Jordan]: Sure! I was thinking that we could add a feature where users can log in using their social media accounts. This can help them easily share content from the app to their social platforms. We could also incorporate some light gamification by awarding badges or points to users when they complete specific actions in the app.

[Alex]: The social login might be feasible, but for the gamification elements, it's important to think through what actions would earn badges or points, and make sure they align with the app's main purpose. We also have to be careful with data privacy issues when integrating social media.

[Jordan]: Good point, Alex. Maybe we can have an opt-in feature for users who are comfortable sharing their data and keep it separate for those who aren't?

[Alex]: That sounds like a balanced approach.

[Teammate 1 and 2 have spoken, waiting for your response.]

**L:** I like that approach but I'd like to learn a little more about how gamification might work. Can either of you tell me a little more about it?

**🤖:** [Jordan]: Absolutely! Gamification involves using game-like elements such as points, badges, and leaderboards in non-game contexts to motivate users. For instance, we could award users points for sharing content from our app to their social media. A leaderboard can show the most active users, which can encourage more engagement.

[Alex]: To add to that, gamification can be effective if done right. However, it's critical that it doesn't feel forced or detract from the primary functions of the app. It's important to not overwhelm the user, and to remain mindful of privacy and data security.

At this point in the meeting, you have a consequential decision to make: whether to go ahead with implementing both social media integration and gamification features, or to take a more conservative approach and only implement social media integration for now.

[Teammate 1 and 2 have spoken, waiting for your decision.]

**L:** Given our time frame for delivery, I would decide to move ahead with social media integration and explore gamification by building out a project plan and testing for that functionality, and potentially incorporate it in future.

**🤖:** [Jordan]: That sounds like a wise decision! We can make a substantial impact with social media integration for now, and thoroughly plan for gamification later on.

[Alex]: I agree. This will give us time to ensure that the social media integration is executed well and we can be thoughtful about how we design and implement gamification.

[Meeting ends]

Feedback on how you ran the meeting:

1. **Positives**:
   - **Inclusion**: You made an effort to include both team members in the conversation, recognizing the importance of hearing both perspectives.
   - **Clarification**: You asked for specifics to ensure you fully understood the proposals and concerns from each team member.
   - **Balance**: You displayed a balanced approach in your decision-making, acknowledging the need to innovate while also considering the complexity and timeline.
   - **Psychological Safety**: You made the environment safe for Alex to speak up. Also, by valuing each person's input, you fostered a sense of respect and openness.
2. **Areas for Improvement**:
   - **Mediation**: You could have stepped in earlier when Jordan was dominating the conversation to ensure that Alex felt encouraged to share more insights.
   - **Action Items**: At the end of the meeting, it would have been good to clearly assign action items and responsibilities to each team member based on the decision made,

## AI as Simulator: Risks

The AI can hallucinate and make up facts about any concept. It may not have enough information about your specific concept or series of concepts and it doesn't know your students' learning level (it has no sense of where your students are). It can also get in a loop in the story so that it loses track of its specific goal (to give the student a chance to practice applying a specific concept). Additionally, no scenario will be effective for all students. Because the AI is creating a story, students may focus on elements of the story and lose track of the general concept; additionally, students may not be familiar with elements of the story, causing extraneous load on their working memory (as they try to both understand the story and apply what they learned given the story).

## AI as Simulator: Guidelines for teachers

Ahead of any practice with the AI, you can let students know of the goal of the exercise: to practice what they learned about [a concept or series of concepts] through role play. You may decide to remind students or explicitly name aspects of the concept they should recall, or you may briefly discuss these with students before the exercise. You can tell students that although they will likely get a similar type of scenario, each will be different. Additionally, because self-monitoring is an essential element of transfer, you can have students write a reflection about the interaction; they can discuss what the AI got right (or wrong) about its feedback. Students can also address why this was (or was not) an effective scenario to practice a skill – that is, did the AI ask a consequential question that challenged the student to apply the specific concept?

## AI as Simulator: Instructions for students

When interacting with the AI-Scenario builder, remember:

**It may simply not work the first time you try it.** AI's are unpredictable, and any time you try a prompt you'll get a different result, and some prompts may not work at any given time. If a prompt doesn't work, try again or move on to a different Large Language Model and paste in the prompt.

**The AI is not a person, but it may feel like one.** Both the scenario is a fiction, and the AI playing a role in the scenario is not a real person responding to you. It doesn't know you or your context.

**You are responsible for your own work.** While the AI can help, it can get things wrong or subtly wrong. You should carefully consider its questions and final feedback and ask yourself: *does the feedback adequately address the concept or sum up my performance in the scenario?*

**It can make "hallucinate" or make things up.** Take every piece of feedback or explanation critically and evaluate the explanation. Check with trusted sources.

**Only share what you are comfortable sharing.** Do not feel compelled to share anything personal. Anything you share may be used as training data for the AI.

Here are a few ways to get the most out of the interaction with the AI Scenario builder:

- **Goal play.** In role playing scenarios, you are given a role to play. But in this scenario, you can play the role but keep the goal of the exercise in mind – to practice what you learned during the scenario. Immerse yourself in the scene and play your role, recalling what you learned and applying it to the challenges in the scene.

- **Give it extensive responses.** The more extensive your answers and explanations the more you can get out of the practice session.

- **Seek clarification**. If you are confused at any point ask questions to help clarify.

- **Try a different scenario**. If you prefer a different scenario try pasting in the prompt again. Because the AI response is randomized the scenario will differ every time you paste in the prompt.

**Share your complete interactions with the AI**. In a paragraph briefly discuss what you learned from using the tool. How well did it work? Did anything surprise you? What are some of your takeaways in working with the AI? What did you learn about your reaction to the scene? What advice or suggestions did it give you? Was the advice helpful?

## AI as Simulators: Build Your Own

To build your own scenario builder, start with the learning goal: For instance, *the goal for this scene is for students to practice their interviewing skills.*

**Goal: Tell the AI what you want it to do and what you don't want it to do**. For instance, *your goal is to give students practice in interviewing a candidate focusing on hypothetical and behavioral questions and follow up questions. Don't play both roles. Wait for the student to respond before moving ahead with the conversation.*

**Role: Tell the AI who it is.** For example, *you will play the role of the candidate and I will play the role .*

**Step-by-step instructions**. For instance, as the interviewer, I *should have to ask questions and the candidate will also get a chance to ask me a question. After 5 interactions, set up a consequential choice for me to make. Then wrap up by telling me how I performed as an interviewer and what I can do better next time.*

**Final Step:** Check your prompt by trying it out. You may want to add more specifics about the concept, and you may want to provide the AI with specifics about your students' learning level. For example, *tailor the interaction for students taking an undergraduate college course and focus on conducting structured interviews*. You can continue to tweak the prompt until it works for you and until you feel that it will work for your students.

## AI as Tool

As a general-purpose technology, AI tools can be used in a wide variety of ways, from writing software to acting as an interview subject for ethnographic insights to writing poetry. This use of an AI, as a tool for extending the amount of work that students can do and the amount we can accomplish, is in many ways the most exciting use of AI. Because many AI uses are highly specific to individual classes and use cases, we encourage instructors to experiment with prompts. Further, educators should share prompts with peers to collectively improve our ability to use AI tools.

## Conclusion

We propose that the advent of AI tools, which facilitate skill development and practice outside the classroom, warrants a re-examination of traditional assignments and classroom practices. As these tools have the potential to aid independent learning and potentially offer personalized engagement, educators can experiment with devoting more in-class time to active discussions, question-and-answer sessions, and collaborative exercises such as 'think-pair-share' sessions. Such changes foster an active learning environment, inviting each student to engage with class concepts, articulate reasoning, and actively construct knowledge. In this scenario, the AI aids with personalized, readily available tutoring and coaching outside of the classroom and the classroom transforms into a hub of systematic engagement. Here, discussion, group work, and reflection on asynchronous learning activities intermingle, while teachers incorporate a variety of questioning techniques (Sherrington, 2020). In this classroom, each student should not only have the opportunity to practice but also actively participate in discussions, creating an inclusive and deeply participatory learning environment.

These approaches are just the start of using AI in class. By sharing the advantages, as well as the risks of these new approaches, educators and students can begin to work together to come up with ways to train students to use AI responsibly, in ways that enhance both their education and life outcomes. The challenges around AI remain significant, as do the opportunities. Only by learning to use these new tools firsthand can students begin to understand their implications, and prepare themselves for a world where AI forms an important part of their lives.

# References

Bjork, E. L., & Bjork, R. A. (2011). Making things hard on yourself, but in a good way: Creating desirable difficulties to enhance learning. *Psychology and the real world: Essays illustrating fundamental contributions to society*, *2*(59-68).

Brown, P. C., Roediger III, H. L., & McDaniel, M. A. (2014). *Make it stick*. Harvard University Press.

Carey, B. (2015). *How we learn: the surprising truth about when, where, and why it happens*. Random House Trade Paperbacks.

Chi, M. T., Roy, M., & Hausmann, R. G. (2008). Observing tutorial dialogues collaboratively: Insights about human tutoring effectiveness from vicarious learning. *Cognitive science*, *32*(2), 301-341.

Chi, M. T., Siler, S. A., Jeong, H., Yamauchi, T., & Hausmann, R. G. (2001). Learning from human tutoring. *Cognitive science*, *25*(4), 471-533.

Deslauriers, L., McCarty, L. S., Miller, K., Callaghan, K., & Kestin, G. (2019). Measuring actual learning versus feeling of learning in response to being actively engaged in the classroom. *Proceedings of the National Academy of Sciences*, *116*(39), 19251-19257.

Di Stefano, G., Pisano, G., & Staats, B. R. (2015). Learning by thinking: How reflection aids performance. In *Academy of Management Proceedings* (Vol. 2015, No. 1, p. 12709). Briarcliff Manor, NY 10510: Academy of Management.

Dunlosky, J., Rawson, K. A., Marsh, E. J., Nathan, M. J., & Willingham, D. T. (2013). Improving Students' Learning with Effective Learning Techniques: Promising Directions from Cognitive and Educational Psychology. *Psychological Science in the Public Interest, 14,* 4–58. http://doi.org/10.1177/1529100612453266

Edmondson, A. C. (2018). *The fearless organization: Creating psychological safety in the workplace for learning, innovation, and growth*. John Wiley & Sons.

Ericsson, K. A., Krampe, R. T., & Tesch-Römer, C. (1993). The role of deliberate practice in the acquisition of expert performance. *Psychological review*, *100*(3), 363.

Ericsson, A., & Pool, R. (2016). *Peak: Secrets from the new science of expertise*. Random House.

Fiorella, L. (2023). Making Sense of Generative Learning. *Educational Psychology Review*, *35*(2), 50.

Fiorella, L., & Mayer, R. E. (2015). *Learning as a generative activity*. Cambridge university press.

Haas, M., & Mortensen, M. (2016). The secrets of great teamwork. *Harvard Business Review*. Retrieved from https://hbr.org/2016/06/the-secrets-of-great-teamwork

Hackman, J. R. (2011). *Collaborative intelligence: Using teams to solve hard problems*. Berrett-Koehler Publishers.

Hill, H. C. (2021). Learning recovery: The research on tutoring, extended school year, and other strategies. Education Week.

Kirschner, P., & Hendrick, C. (2020). *How learning happens: Seminal works in educational psychology and what they mean in practice*. Routledge.

Kirschner, P., & Neelen, M. (2018). No Feedback, No Learning. 3-Star Learning Experiences. https://3starlearningexperiences.wordpress.com/2018/06/05/no-feedback-no-learning/

Klein, G. (2007). Performing a project premortem. *Harvard business review*, *85*(9), 18-19.

Kraft, M., Schueler, B., Loeb, S., & Robinson, C. (February, 2021). Accelerating Student Learning with High- Dosage Tutoring. Annenberg Institute for School Reform at Brown University.

Kross, E., & Ayduk, O. (2017). Self-distancing: Theory, research, and current directions. In *Advances in experimental social psychology* (Vol. 55, pp. 81-136). Academic Press.

Mccrea, P. (2023). Developing expert teaching: A practical guide to designing effective professional development, for others and ourselves (High impact teaching).

Metcalfe, J. (2017). Learning from errors. *Annual review of psychology*, *68*, 465-489.

Mitchell, D. J., Edward Russo, J., & Pennington, N. (1989). Back to the future: Temporal perspective in the explanation of events. *Journal of Behavioral Decision Making*, *2*(1), 25-38.

Mollick, E.R. & Mollick, L. (2022). New Modes of Learning Enabled by AI Chatbots: Three Methods and Assignments (December 13, 2022).

Mollick, E. R., & Mollick, L. (2023). Using AI to implement effective teaching strategies in classrooms: Five strategies, including prompts. *Including Prompts (March 17, 2023)*.

OpenAI. (2023). GPT-4 technical report. Retrieved from https://openai.com/research/gpt-4/

Perkins, D. N., & Salomon, G. (1992). Transfer of learning. *International encyclopedia of education*, *2*, 6452-6457.

Rawson, K. A., Thomas, R. C., & Jacoby, L. L. (2015). The power of examples: Illustrative examples enhance conceptual learning of declarative concepts. *Educational Psychology Review*, *27*, 483-504.

Roscoe, R. D., & Chi, M. T. (2007). Understanding tutor learning: Knowledge-building and knowledge-telling in peer tutors' explanations and questions. *Review of educational research*, *77*(4), 534-574.

Schön, D. A. (1987). *Educating the reflective practitioner: Toward a new design for teaching and learning in the professions*. Jossey-Bass.

Seligman, M. E., Railton, P., Baumeister, R. F., & Sripada, C. (2013). Navigating into the future or driven by the past. *Perspectives on psychological science*, *8*(2), 119-141.

U.S. Department of Education, Office of Educational Technology, Artificial Intelligence and Future of Teaching and Learning: Insights and Recommendations, Washington, DC, 2023.

Walton Family Foundation. (2023). Teachers and Students Embrace ChatGTP For Education. Retrieved May 27, 2023, from https://www.waltonfamilyfoundation.org/chatgpt-report

Wiggins, G. (2015). Seven keys to effective feedback. ACSD. https://www.ascd.org/el/articles/seven-keys-to-effectivefeedback

Willingham, D. T. (2021). *Why don't students like school?: A cognitive scientist answers questions about how the mind works and what it means for the classroom*. John Wiley & Sons.

Willingham, D. T. (2023). *Outsmart Your Brain: Why Learning is Hard and How You Can Make It Easy*. Simon and Schuster.

Wiliam, D. (2011). What is assessment for learning?. *Studies in educational evaluation*, *37*(1), 3-14.

Wiliam, D., & Leahy, S. (2016). *Embedding formative assessment*. Hawker Brownlow Education.

**Appendix: Large Language Models and Prompt Compatibility**

| Approach | OpenAI ChatGPT 4 | OpenAI ChatGPT 3.5 | Microsoft's Bing in Creative Mode | Anthropic's Claude | Google's Bard |
|---|---|---|---|---|---|
| **Increasing Knowledge** AI-Tutor | yes | no | yes | no | no |
| **Increasing Metacognition:** AI Team Reflection Coach | yes | no | yes | yes | no |
| **Increasing Metacognition** AI Coach: Team Premortem | yes | no | yes | no | no |
| **Providing Feedback:** AI Mentor | yes | no | yes | no | no |
| **Building Collective Intelligence:** AI Teammate | yes | sometimes | yes | sometimes | no |
| **Increasing Fluency:** AI Student | yes | yes | yes | yes | yes |
| **Practice:** AI Simulator | yes | sometimes | yes | sometimes | no |

*A chart for prompts that work with Large Language Models in this paper. Note: subject to change, as the models change.*